\documentclass[11pt]{article}

\usepackage{amssymb}
\usepackage{amsmath}

\newtheorem{theorem}{Theorem}

\newtheorem{corollary}[theorem]{Corollary}

\newtheorem{proposition}{Proposition}

\begin{document}


\title
{Inverse Boundary Value Problem by Partial data for
the Neumann-to-Dirichlet-map in two dimensions}

\author{
O.~Yu.~Imanuvilov, \thanks{ Department of Mathematics, Colorado
State University, 101 Weber Building, Fort Collins, CO 80523-1874,
USA e-mail: oleg@math.colostate.edu}\, Gunther~Uhlmann,\thanks{
Department mathematics, UC Irvine, Irvine CA 92697 Department of
Mathematics, University of Washington, Seattle, WA 98195 USA e-mail:
gunther@math.washington.edu The second author partly supported by
NSF and a Walker Family Endowed Professorship}\, M.~Yamamoto\thanks{
Department of Mathematical Sciences, University of Tokyo, Komaba,
Meguro, Tokyo 153, Japan e-mail: myama@ms.u-tokyo.ac.jp}}

\date{}

\maketitle

\begin{abstract}
For the two dimensional Schr\"odinger equation in a bounded domain,
we prove uniqueness of determination of potentials in
$W^1_p(\Omega),\,\, p>2$ in the case where we apply all possible
Neumann data supported on an arbitrarily non-empty open set
$\widetilde\Gamma$ of the boundary and observe the corresponding
Dirichlet data on $\widetilde{\Gamma}$. An immediate consequence is
that one can uniquely determine a conductivity in $W^3_p(\Omega)$
with $p>2$ by measuring the voltage on an open subset of the
boundary corresponding to current supported in the same set.
\end{abstract}


\section{\bf Introduction}

Let $\Omega\subset \Bbb R^2$ be a bounded domain with smooth
boundary $\partial\Omega$ and  let $\nu=(\nu_1, \nu_2)$ be the unit
outer normal to $\partial\Omega$ and let
$\frac{\partial}{\partial\nu} = \nabla\cdot\nu$.

In this domain we consider the Schr{\"o}dinger equation with a
potential $q$:
\begin{equation}
L_q(x,D)u=(\Delta +q)u=0\quad\mbox{in}\,\,\Omega.
\end{equation}

Let $\widetilde \Gamma$ be a non-empty arbitrary fixed relatively
open subset of $\partial\Omega.$ Consider the Neumann-to-Dirichlet map $N_{q,\widetilde\Gamma}$  with partial  data on $\tilde \Gamma$
defined by
\begin{equation}\label{popo}
N_{q,\widetilde\Gamma}: f\rightarrow u\vert_{\widetilde\Gamma},
\end{equation} where
\begin{equation}\label{lopo}
(\Delta+q)u=0\quad\mbox{in}\,\,\Omega,\,\, \frac{\partial
u}{\partial \nu}\vert_{\partial\Omega\setminus \widetilde\Gamma}=0,
\,\, \frac{\partial u}{\partial \nu}\vert_{\widetilde\Gamma}=f
\end{equation} with domain $D(N_{q,\widetilde\Gamma})\subset L^2(\widetilde \Gamma).$
Without loss of generality
we may assume that $\partial\Omega\setminus\widetilde \Gamma$ contains
a non-empty open set.
By uniqueness of the Cauchy problem for the Sch\"odinger equation
the operator $N_{q,\widetilde\Gamma}$ is well defined
since the problem (\ref{lopo}) has at most one solution
for each $f\in L^2(\widetilde \Gamma).$
Thanks to the Fredholm alternative, we see that
$D(N_{q,\widetilde\Gamma})=\overline{D(N_{q,\widetilde\Gamma})}$ and
$L^2(\widetilde{\Gamma}) \setminus D(N_{q,\widetilde\Gamma})$ is finite
dimensional for any potential $q$ in $W^1_2(\Omega).$

The goal of this article is to prove uniqueness of the determination
of the potential $q$ from the Neumann-to-Dirichlet map
$N_{q,\widetilde\Gamma}$ given by (\ref{popo}) for
arbitrary subboundary $\widetilde \Gamma$. More precisely,
we consider all Neumann data supported on an
arbitrarily fixed subboundary $\widetilde{\Gamma}$ as input and we
observe the Dirichlet data only on the same subboundary
$\widetilde\Gamma$. This map arises in electrical impedance tomography (EIT)
where one attempts to determine the electrical conductivity
of a medium by inputting voltages and measuring current at
the boundary. After transforming
(1) to the conductivity equation, we can interpret $u\vert_{\widetilde\Gamma}$
and $\frac{\partial u}{\partial\nu}_{\widetilde\Gamma}$ respectively
as the voltage and the multiple of the current by values of the
surface conductivity.
In practice, we can realize such inputs and outputs by applying
current to electrodes on the
boundary and observing the corresponding voltages. The
current inputs are modeled by the Neumann boundary data
$\frac{\partial u}{\partial \nu}$ and the observation data is
modeled by Dirichlet data. See e.g., Cheney, Issacson and Newell \cite{CIN}
for applications to medical imaging of EIT. Moreover it is very desirable to
 restrict the supports of the current inputs as small as
possible. To the authors' best knowledge there are few works on the
uniqueness by such a ''Neumann-to-Dirichlet map" with partial data.
In Astala, P\"aiv\"arinta and Lassas \cite{APL}, the authors consider both the
Dirichlet-to-Neumann map and the Neumann-to-Dirichlet map on an
arbitrarily subboundary to establish the uniqueness of an anisotropic
conductivity modulo the group of diffeomorphisms which is the identity on the
boundary where the measurements take place.

The case where the measurements are given by the Dirichlet-to-Neumann map has been extensively
studied in the literature. This map is defined in the case of partial data by
$$
\Lambda_{q,\widetilde \Gamma}: g\rightarrow \frac{\partial
u}{\partial\nu}\vert_{\widetilde\Gamma};\quad
 (\Delta+q)u=0\quad\mbox{in}\,\,\Omega,\,\,
u\vert_{\partial\Omega\setminus \widetilde\Gamma}=0 ,\quad
u\vert_{\widetilde\Gamma}=g.
$$

We give some references but the list is not at all complete. In the case of
full data $\widetilde{\Gamma} =
\partial\Omega$, this inverse problem was formulated by Calder\'on
\cite{C}. In the two dimensional case, given a Dirichlet-to-Neumann
map $\Lambda_{q,\widetilde \Gamma}$ on an arbitrary subbondary
$\widetilde \Gamma$, uniqueness is proved under the assumption $q\in
C^{2+\alpha}(\overline\Omega)$ by Imanuvilov, Uhlmann and Yamamoto
\cite{IUY} and for the uniqueness for potentials $q\in
W^1_p(\Omega), p>2$ see Imanuvilov and Yamamoto \cite{IY2}. For
other uniqueness results by the Dirichlet-to-Neumann map on an
arbitrary subboundary $\widetilde{\Gamma}$, we can refer also to
Imanuvilov, Uhlmann and Yamamoto \cite{IUY2}, \cite{IUY1}. Also see
Imanuvilov and Yamamoto \cite{IY1} for uniqueness results for
elliptic systems.  In Guillarmou and Tzou \cite{GZ}, the result of
\cite{IUY} was extended on Riemmannian surfaces. In particular, for
uniqueness in determining a two-dimensional potential with full
data: $\Lambda_{q,\partial\Omega}$, we refer to Blasten \cite{EB},
Bukhgeim \cite{Bu} and, Sun and Uhlmann \cite{SuU}, and for systems
in Albin, Guillarmou, Tzou and Uhlmann \cite{AGTU} and Novikov and
Santacesaria \cite{Nov-San1}. For the case of full data, in
\cite{EB} and \cite{IY2},  it was shown that
$\Lambda_{q,\partial\Omega}$ uniquely determines $q$  in the class
piecewise $W^1_p(\Omega)$ with $p>2$ and
$C^\alpha(\overline\Omega),\alpha>0$, respectively. As for the
related problem of recovery of the conductivity in EIT, Astala and
P\"aiv\"arinta \cite{AP} proved uniqueness for conductivities in
$L^\infty(\Omega),$ improving the results of Nachman \cite{N} and
Brown and Uhlmann \cite{BrU}. Moreover for the case of dimensions $n
\ge 3$ with the full data Sylvester and Uhlmann \cite{SU} proved the
uniqueness of recovery of a conductivity in $C^2(\overline\Omega)$,
and later the regularity assumption was improved (see, e.g., Brown
and Torres \cite{BT}, P\"aiv\"arinta, Panchenko and Uhlmann
\cite{PPU} and Haberman and Tataru \cite{HT}). The case when
voltages are applied and current is measured on different subsets
was studied in dimensions greater than three in Bukhgeim and Uhlmann
\cite{BuU}, Kenig, Sj\"ostrand and Uhlmann \cite{KSU} and in
Imanuvilov, Uhlmann and Yamamoto \cite{IUY3} for the two-dimensional
case.

Our main result is as follows
\begin{theorem}\label{001}
Let  $q_1,q_2\in W^1_p(\Omega)$ for some $p>2.$ If $
D(N_{q_1,\widetilde\Gamma})\subset D(N_{q_2,\widetilde\Gamma})$ and
$N_{q_1,\widetilde \Gamma}(f)= N_{q_2,\widetilde\Gamma}(f)$ for each $f$
from $ D(N_{q_1,\widetilde\Gamma})$, then $q_1=q_2$ in $\Omega$.
\end{theorem}

%

Notice that Theorem \ref{001} does not assume that $\Omega$ is
simply connected.  An interesting inverse problem is whether one can
determine the potential in a domain with holes by measuring
$N_{q,\widetilde\Gamma}$
only on some open set $\widetilde{\Gamma}$ in the outer subboundary.

Let $\Omega,G$ be bounded domains in $\Bbb R^2$ with smooth
boundaries such that $\overline G\subset \Omega.$ Let
$\widetilde{\Gamma} \subset \partial\Omega$ be an open set and $q\in
W^1_p(\Omega\setminus \overline{G})$ with some $p > 2$.
Consider
the following  Neumann-to-Dirichlet map:
\begin{equation}\nonumber
\widetilde N_{q,\widetilde\Gamma}: f\rightarrow u\vert_{\widetilde\Gamma},
\end{equation}
where
\begin{equation}
\thinspace u \in H^1(\Omega\setminus \overline{G}), \quad
(\Delta+q)u=0\,\,\mbox{in}\,\,\Omega\setminus\overline G, \quad
\frac{\partial u}{\partial\nu}\vert_{\partial G\cup
(\partial\Omega\setminus \widetilde\Gamma)}=0, \quad \frac{\partial
u}{\partial\nu}\vert_{ \widetilde\Gamma}=f. \nonumber
\end{equation}
Then we can directly derive the following from Theorem 1.
\begin{corollary}\label{coro2}
Let $q_1, q_2 \in W^1_p(\Omega\setminus\overline{G})$ with some $p >
2$. If $ D(\widetilde N_{q_1,\widetilde\Gamma})\subset D(\widetilde
N_{q_2,\widetilde\Gamma})$ and $\widetilde N_{q_1,\widetilde
\Gamma}(f)= \widetilde N_{q_2,\widetilde\Gamma}(f)$ for each $f$
from $ D(\widetilde N_{q_1,\widetilde\Gamma})$, then $q_1=q_2$ in
$\Omega\setminus \overline{G}$.
\end{corollary}

For the case of EIT, if the conductivities are known on
$\widetilde\Gamma$, then we can apply our theorem to prove
uniqueness of the determination of conductivities in $W^3_p(\Omega), p>2$
from the Neumann-to-Dirichlet map.
\\
\vspace{0.3cm}
\\

The remainder of the paper is devoted to the proof of the theorem
\ref{001}. The main technique is the construction of complex
geometrical optics solutions whose Neumann data vanish on the
complement of $\widetilde \Gamma$.

Throughout the article, we use the following notations.

{\bf Notations.}
We set $\Gamma_0=\partial\Omega\setminus\overline{\widetilde\Gamma},$
$i=\sqrt{-1}$, $x_1, x_2 \in {\Bbb R}^1$, $z=x_1+ix_2$,
$\overline{z}$ denotes the complex conjugate of $z \in \Bbb C$. We
identify $x = (x_1,x_2) \in {\Bbb R}^2$ with $z = x_1 +ix_2 \in
{\Bbb C}$ and $\xi=(\xi_1,\xi_2)$ with $\zeta=\xi_1+i\xi_2$.
$\partial_z = \frac 12(\partial_{x_1}-i\partial_{x_2})$,
$\partial_{\overline z}= \frac12(\partial_{x_1}+i\partial_{x_2})$,
$D = \left( \frac{1}{i}\partial_{x_1},
\frac{1}{i}\partial_{x_2}\right),$ $\partial_\zeta=\frac
12(\partial_{\xi_1}-i\partial_{\xi_2}), \partial_{\overline \zeta}=
\frac12(\partial_{\xi_1}+i\partial_{\xi_2}).$ Denote by
$B(x,\delta)$ a ball centered at $x$ of radius $\delta.$ For a
normed space $X$, by $o_X(\frac{1}{\tau^\kappa})$ we denote a
function $f(\tau,\cdot)$ such that $ \Vert
f(\tau,\cdot)\Vert_X=o(\frac{1}{\tau^\kappa})\quad \mbox{as}
\,\,\vert \tau\vert\rightarrow +\infty$.  The tangential derivative
on the boundary is given by
$\partial_{\vec\tau}=\nu_2\frac{\partial}{\partial x_1}
-\nu_1\frac{\partial}{\partial x_2}$, where $\nu=(\nu_1, \nu_2)$ is
the unit outer normal to $\partial\Omega$.
 The operators
$\partial_{z}^{-1}$ and $\partial_{\overline z}^{-1}$ are given by
$$
\partial_{\overline z}^{-1}g=-\frac1\pi\int_\Omega
\frac{g(\zeta,\overline\zeta)}{\zeta-z} d\xi_2d\xi_1,\quad
\partial_{ z}^{-1}g
= \overline{\partial^{-1}_{\overline{z}}\overline{g}}.
$$
We call $b(z)$ antiholomorphic if $b(\overline{z})$ is holomorphic.
In the Sobolev space $W_2^1(\Omega)$ we introduce the following norm
$$
\Vert u\Vert_{W_2^{1,\tau}(\Omega)} = (\Vert
u\Vert_{W_2^1(\Omega)}^2 + \vert \tau\vert^2\Vert
u\Vert^2_{L^2(\Omega)})^{\frac{1}{2}}.
$$
\vspace{0.3cm}

\section{Proof of Theorem 1}

Let $\Phi=\varphi+i\psi$ be a holomorphic function in
$\Omega$ such that $\varphi, \psi$ are real-valued and
\begin{equation}\label{1} \Phi\in C^2(\overline\Omega),\quad
\mbox{Im}\, \Phi\vert_{\Gamma_0^*}=0, \quad \Gamma_0\subset\subset
\Gamma_0^*,
\end{equation}
where $\Gamma_0^*$ is some open set in $\partial\Omega.$ Denote by
$\mathcal H$ the set of the critical points of the function $\Phi.$
Assume that
\begin{equation}\label{22}
\mathcal H\ne \emptyset,\quad\partial^2_z\Phi(z)\ne 0,\quad \forall
z\in \mathcal H,\quad \mathcal
H\cap\overline{\widetilde\Gamma}=\emptyset
\end{equation}
 and
\begin{equation}\label{kk}
\int_{\mathcal J}1d\sigma =0,\quad \mathcal J=\{x; \thinspace
\partial_{\vec \tau}\psi(x)=0,x\in \partial\Omega\setminus\Gamma_0^*\}.
\end{equation}

Let $\Omega_1$ be a bounded domain in $\Bbb R^2$ such that
$\Omega\subset\subset \Omega_1$ and  $\mathcal C$ be some smooth
complex-valued function in $\Omega$ such that
\begin{equation}\label{nono}
2\frac{\partial \mathcal C}{\partial z} =C_1(x)+iC_2(x)\quad
\mbox{in}\,\,\Omega_1,
\end{equation}
where $C_1,C_2$ are smooth real-valued functions in $\Omega$ such
that
\begin{equation}\label{zopaW}
\frac{\partial C_1}{\partial x_1}+\frac{\partial C_2}{\partial
x_2}=1\quad \mbox{in}\quad\Omega_1.
\end{equation}

The following proposition is proved as Proposition 2.5 in \cite{IUY1}.
\begin{proposition}\label{Theorem 2.1}
Suppose that $q\in L^\infty(\Omega)$, the function $\Phi$ satisfies
(\ref{1}), (\ref{22}), and the function $\mathcal C$ satisfies
(\ref{nono}), (\ref{zopaW}) and $\widetilde v\in W_1^2(\Omega)$ .
Then there exist $\tau_0$ and $C(N)$ independent of $\widetilde v$
and $\tau$ such that

\begin{eqnarray}\label{xxx} \frac{N}{2}\Vert 2\partial_{\overline z}
\widetilde v e^{\tau\varphi+N \mathcal C}\Vert^2 _{L^2(\Omega)}
+\tau\Vert \widetilde v e^{\tau\varphi+N\mathcal
C}\Vert^2_{L^2(\Omega)} +\Vert \widetilde v e^{\tau\varphi+N\mathcal
C}\Vert^2
_{W_2^1(\Omega)}+\tau^2\Vert\vert\frac{\partial\Phi}{\partial z}
\vert \widetilde v e^{\tau\varphi+N\mathcal C}\Vert^2_{L^2(\Omega)}
\nonumber\\
\le \Vert L_q(x,D)\widetilde v e^{\tau\varphi+N \mathcal C}\Vert^2
_{L^2(\Omega)} + C(N)\tau\Vert (\widetilde v
e^{\tau\varphi+N\mathcal C} ,\frac{\partial\widetilde
v}{\partial\nu}e^{\tau\varphi+N\mathcal C}
)\Vert^2_{W_2^{1,\tau}(\partial\Omega)\times L^2(\partial\Omega)}
\end{eqnarray}
for all $\tau > \tau_0(N)$ and all positive $N\ge 1.$
\end{proposition}
Let $\widetilde v\in W_2^2(\Omega)$ satisfy
$$
L_q(x,D)\widetilde v=f\quad \mbox{in}\,\,\Omega,\quad \frac{\partial
\widetilde v}{\partial\nu}\vert_{\Gamma^*_0}=0.
$$
Using Proposition \ref{Theorem 2.1}, we can show the following.
\begin{proposition}\label{Theorem 2.2}
Suppose that $\Phi$ satisfies (\ref{1}), (\ref{22}) and $q\in
L^\infty(\Omega) .$ Then there exist $\tau_0$ and $C$ independent of
$\widetilde v$ and $\tau$ such that
\begin{eqnarray}\label{Xxxx1}
\tau\Vert \widetilde v e^{\tau\varphi}\Vert^2_{L^2(\Omega)} +\Vert
\widetilde v e^{\tau\varphi}\Vert^2
_{W_2^1(\Omega)}+\tau^2\Vert\vert\frac{\partial\Phi}{\partial z}
\vert
\widetilde v e^{\tau\varphi}\Vert^2_{L^2(\Omega)} \nonumber\\
\le C\left(\Vert L_q(x,D)\widetilde v e^{\tau\varphi}\Vert^2
_{L^2(\Omega)} + \tau\Vert (\widetilde v
e^{\tau\varphi},\frac{\partial \widetilde v
e^{\tau\varphi}}{\partial\nu})\Vert^2_{W_2^{1,\tau}(\widetilde
\Gamma)\times L^2(\widetilde\Gamma)}\right)
\end{eqnarray}
for all $\tau > \tau_0$ and for all $\widetilde v\in H^2(\Omega).$
\end{proposition}

{\bf Proof.} Let $\{e_j\}_{j=1}^M$ be a partition of unity such that
$e_j\in C^\infty_0(B(x_j,\delta))$ where $x_j$ are some points in
$\Omega,$
$$
\sum_{j=1}^Me_j(x)=1\quad \mbox{on}\,\,\Omega,\quad \frac{\partial
e_j}{\partial\nu}\vert_{\Gamma_0^*}=0\quad\forall j\in\{1,\dots, M\},
$$
and $\delta$ be a small positive number such that $B(x_j,\delta)\cap
\Gamma_0\ne\emptyset$ implies $B(x_j,\delta)\cap
\partial\Omega\subset \Gamma_0^*$. Denote $w_j=e_j\widetilde v.$ Let
$\mbox{supp}\, w_j\cap (\partial\Omega\setminus\widetilde
\Gamma)=\emptyset$. Then Proposition \ref{Theorem 2.1} implies that
there exists $\tau_0$ such that for all  $\tau\ge\tau_0$
\begin{eqnarray}\label{bp}
\frac{N}{2}\Vert 2\partial_{\overline z}w_j e^{\tau\varphi}
e^{N \mathcal C}\Vert^2 _{L^2(\Omega)}\nonumber\\
+ \tau\Vert w_j e^{\tau\varphi+N \mathcal C}\Vert^2_{L^2(\Omega)}
+\Vert w_j e^{\tau\varphi+N \mathcal C}\Vert^2
_{H^1(\Omega)}+\tau^2\Vert\vert\frac{\partial\Phi}{\partial z} \vert
w_j e^{\tau\varphi+N \mathcal C}\Vert^2_{L^2(\Omega)} \nonumber\\
\le C\Vert L_q(x,D)w_j e^{\tau\varphi+N \mathcal
C}\Vert^2_{L^2(\Omega)}+C(N)\Vert ( w_j
e^{\tau\varphi},\frac{\partial w_j
e^{\tau\varphi}}{\partial\nu})\Vert^2_{W_2^{1,\tau}(\widetilde
\Gamma)\times L^2(\widetilde \Gamma)}.
\end{eqnarray}
Next let $\mbox{supp}\, w_j\cap
(\partial\Omega\setminus\widetilde\Gamma) \ne\emptyset$.  We can not
apply directly the  Carleman estimate (\ref{Xxxx1}) in this case,
since the function $w_j$ may not satisfy the zero Dirichlet boundary
condition. To overcame this difficulty we construct an extension.
Without loss of generality, using if necessary a conformal
transformation, we can assume that $\mbox{supp}\, w_j\cap
\Omega\subset \{x_2>0\}$ and $\mbox{supp}\, w_j\cap
\partial\Omega\subset \{x_2=0\}.$
Then using the extension $w_j(x_1,x_2)=w_j(x_1,-x_2),
q(x_1,x_2)=q(x_1,-x_2)$ and $\varphi(x_1,x_2)=\varphi(x_1,-x_2)$, we
apply Proposition \ref{Theorem 2.1} to the operator $L_q(x,D)$ in
$\mathcal O=\mbox{supp}\, e_j\cup \{x\vert (x_1,-x_2)\in
\mbox{supp}\, e_j\}.$ We have the same estimate (\ref{bp}).
Therefore for all $\tau\ge \tau_0$
\begin{eqnarray}\label{xxx1}
\Vert\widetilde v\Vert^2_* := \tau\Vert \widetilde v
e^{\tau\varphi+N\mathcal C}\Vert^2_{L^2(\Omega)}
+\Vert \widetilde v e^{\tau\varphi+N\mathcal C}\Vert^2
_{H^1(\Omega)}+\tau^2\Vert\vert\frac{\partial\Phi}{\partial z} \vert
\widetilde v e^{\tau\varphi+N\mathcal C}\Vert^2_{L^2(\Omega)} \nonumber\\
\le \sum_{j=1}^M\Vert\widetilde  ve_j\Vert^2_* \le
C\sum_{j=1}^M\Vert L_q(x,D)w_j e^{\tau\varphi+N \mathcal
C}\Vert^2_{L^2(\Omega)}+C(N)\Vert (w_j
e^{\tau\varphi},\frac{\partial w_j
e^{\tau\varphi}}{\partial\nu})\Vert^2_{W_2^{1,\tau}(\widetilde\Gamma)\times
L^2(\widetilde\Gamma)}\nonumber\\
\le C \sum_{j=1}^M(\Vert \Delta e_j\widetilde v e^{\tau\varphi+N
\mathcal C}\Vert^2_{L^2(\Omega)}+\Vert2\partial_z e_j
\partial_{\overline z}\widetilde v
e^{\tau\varphi+N \mathcal C}\Vert^2_{L^2(\Omega)}-
N\Vert
\partial_{\overline z}(\widetilde v e_j)e^{\tau\varphi+N \mathcal C}
\Vert^2_{L^2(\Omega)}\nonumber\\
+ \Vert L_q(x,D)\widetilde v e^{\tau\varphi+N \mathcal
C}\Vert^2_{L^2(\Omega)})+C(N)\Vert (\widetilde v
e^{\tau\varphi},\frac{\partial \widetilde
ve^{\tau\varphi}}{\partial\nu})\Vert^2_{W_2^{1,\tau}(\widetilde\Gamma)\times
L^2(\widetilde\Gamma)}.
\end{eqnarray}
Fixing the parameter $N$ sufficiently large, we obtain from
(\ref{xxx1})
\begin{eqnarray}\label{xxx2} \Vert\widetilde v\Vert^2_*
\le C(N)\left(\Vert \widetilde v  e^{\tau\varphi+N \mathcal
C}\Vert^2_{L^2(\Omega)}+ \Vert L_q(x,D)\widetilde v e^{\tau\varphi+N
\mathcal C}\Vert^2_{L^2(\Omega)}+\Vert (\widetilde v
e^{\tau\varphi},\frac{\partial \widetilde
ve^{\tau\varphi}}{\partial\nu})\Vert^2_{W_2^{1,\tau}(\widetilde\Gamma)\times
L^2(\widetilde\Gamma)}\right).
\end{eqnarray}
The first term on the right-hand side of (\ref{xxx2}) can be
absorbed into the left-hand side for all sufficiently large $\tau.$
Since $N$ and $\mathcal C$ are independent of $\tau$, the proof of
the proposition is finished.$\square$

The Carleman estimate (\ref{Xxxx1}) implies the existence of solutions
to the following boundary value problem.
\begin{proposition}\label{Theorem 2.3}
There exists a constant $\tau_0$ such that for $\vert \tau\vert\ge
\tau_0$ and any $f\in L^2(\Omega),r\in W_2^\frac 12 (\Gamma_0^*)$,
there exists a solution to the boundary value problem
\begin{equation}\label{lola}
L_q(x,D)u =fe^{\tau\varphi}\quad\mbox{in}\,\,\Omega, \quad
\frac{\partial u}{\partial \nu}\vert_{\Gamma_0}=re^{\tau\varphi}
\end{equation}
 such that
\begin{equation}\label{2}
 \Vert u\Vert_{W_2^{1,\tau}(\Omega)}/\root\of{\vert\tau\vert}
\le C(\Vert  f\Vert_{L^2(\Omega)}+\vert\tau\vert^\frac 14 \Vert
r\Vert_{L^2 (\Gamma_0)}+\Vert r\Vert_{W_2^{\frac 12}
 (\Gamma^*_0)}).
 \end{equation}
The constant $C$ is independent of $\tau.$
\end{proposition}

The proof of this proposition uses standard duality arguments,
see  e.g., \cite{IUY}.
\\
\vspace{0.3cm}

We define two other operators:
\begin{equation}\label{anna}
\mathcal R_{\tau}g = \frac 12e^{\tau(\Phi - \overline{\Phi})}
\partial_{\overline z}^{-1}(g
e^{\tau(\overline{\Phi}-\Phi)}),\,\, \widetilde {\mathcal R}_{\tau}g
= \frac 12 e^{\tau(\overline {\Phi}-{\Phi})}
\partial_{ z}^{-1}(ge^{\tau( {\Phi}
-\overline {\Phi})}).
\end{equation}
Observe that \begin{equation}\label{begemot}
2\frac{\partial}{\partial z}(e^{\tau\Phi}\widetilde {\mathcal
R}_{\tau}g)=ge^{\tau\Phi},\quad 2\frac{\partial}{\partial \overline
z}(e^{\tau\overline\Phi} {\mathcal
R}_{\tau}g)=ge^{\tau\overline\Phi} \quad \forall g\in L^2(\Omega).
\end{equation}
Let $a\in C^6(\overline\Omega)$ be some holomorphic function on
$\Omega$ such that
\begin{equation}\label{LL}
\mbox{Im}\, a\vert_{\Gamma^*_0}=0 ,\quad
\lim_{z\rightarrow \hat z}a(z)/\vert z-\hat z\vert^{100}=0, \quad
\forall \hat z\in \mathcal H\cap\Gamma^*_0.
\end{equation}

Moreover, for some $\widetilde x\in \mathcal H$, we assume that
\begin{equation}\label{begemot}
a(\widetilde x) \ne 0\quad \mbox{and}\quad a(x) = 0,\quad\forall x
\in \mathcal H\setminus \{\widetilde x\}. \end{equation}

The existence of such a function is proved in Proposition 9 of
\cite{IY1}. Let polynomials $M_{1}(z)$ and $M_{3}(\overline z)$
satisfy
\begin{equation}\label{begemot2}
(\partial^{-1}_{\overline z}q_{1} -M_{1})(\widetilde x)=0, \quad \quad
(\partial^{-1}_{z}q_{1} - M_{3})(\widetilde x)= 0.
\end{equation}

The holomorphic function $a_1$ and the antiholomorphic function
$b_1$ are defined by formulae
$a_1(z)=a_{1,1}(z)+a_{1,2}(z)+a_{1,3}(z)$ and $b_1(\overline
z)=b_{1,1}(\overline z)+b_{1,2}(\overline z) +b_{1,3}(\overline z)$
where  $a_{1,1},b_{1,1}\in C^1(\overline \Omega)$ and
\begin{equation}\label{monica}
i\frac{\partial\psi}{\partial\nu}
a_{1,1}(z)-i\frac{\partial\psi}{\partial\nu} b_{1,1}(\overline
z)=-\frac{\partial (a+\overline
a)}{\partial\nu}+i\frac{\partial\psi}{\partial\nu}
\frac{a(\partial^{-1}_{\overline z}
q_{1}-M_{1})}{4\partial_z\Phi}-i\frac{\partial\psi}{\partial\nu}
\frac{\overline{a}(\partial^{-1}_{z} q_{1}-M_{3})}
{4\partial_{\overline z}\overline\Phi}
\quad\mbox{on}\,\,\Gamma_0^*
\end{equation}
and $a_{1,2}(z,\tau),b_{1,2}(\overline z,\tau)\in C^1(\overline \Omega)$
for each $\tau$ are holomorphic and antiholomorphic functions  such
that
$$
b_{1,2}(\overline z,\tau)=-\frac{1}{8\pi}\int_{\partial\Omega}
\frac{(\nu_1-i\nu_2) a(\partial^{-1}_{\overline
\zeta}q_{1}-M_{1})e^{\tau(\Phi-\overline\Phi)}}{(\overline\zeta-\overline
z)\partial_\zeta\Phi} d\sigma
$$
and
$$
a_{1,2}(z,\tau)=-\frac{1}{8\pi}\int_{\partial\Omega}
\frac{(\nu_1+i\nu_2)\overline{a}(\partial_\zeta^{-1}q_{1}-M_{3})
e^{\tau(\overline\Phi-\Phi)}}{(\zeta-z)\partial
_{\overline\zeta}\overline\Phi} d\sigma.
$$
Here the denominators of the integrands vanish in ${\cal H} \cap
\Gamma^*_0$, but thanks to the second condition in (\ref{LL})
integrability is guaranteed. We represent the functions
$a_{1,2}(z,\tau),b_{1,2}(\overline z,\tau)$ in the form
$$
a_{1,2}(z,\tau)=a_{1,2,1}(z)+a_{1,2,2}(z,\tau),\quad
b_{1,2}(\overline z,\tau)=b_{1,2,1}(\overline z)
+ b_{1,2,2}(\overline z,\tau),
$$
where
$$
b_{1,2,1}(\overline z)=-\frac{1}{8\pi}\int_{\Gamma_0^*}
\frac{(\nu_1-i\nu_2) a(\partial^{-1}_{\overline
\zeta}q_{1}-M_{1})}{(\overline\zeta-\overline z)\partial_\zeta\Phi}
d\sigma ,\quad a_{1,2,1}(z)=-\frac{1}{8\pi}\int_{\Gamma_0^*}
\frac{(\nu_1+i\nu_2)\overline{a}(\partial_\zeta^{-1}q_{1}-M_{3})
}{(\zeta-z)\partial _{\overline\zeta}\overline\Phi} d\sigma.
$$

By (\ref{LL}), the functions $b_{1,2,1},a_{1,2,1}$ belong to
$C^1(\overline \Omega).$
By (\ref{kk}) we have
$$
\Vert b_{1,2,1}(\cdot,\tau)\Vert_{L^2(\Omega)}+\Vert
a_{1,2,1}(\cdot,\tau)\Vert_{L^2(\Omega)}\rightarrow 0\quad\mbox{as}
\,\,\tau\rightarrow +\infty .
$$

Finally  $a_{1,3}(z,\tau),b_{1,3}(\overline z,\tau)\in W^1_2(
\Omega)$ for each $\tau$ are holomorphic and antiholomorphic
functions respectively such that
\begin{eqnarray}\label{dom}
i\frac{\partial\psi}{\partial\nu}
a_{1,3}(z,\tau)-i\frac{\partial\psi}{\partial\nu} b_{1,3}(\overline
z,\tau)=\frac{i}{2\pi}\frac{\partial\psi}{\partial\nu}\int_{\Omega}
\partial_\zeta\left (\frac{ a(\partial^{-1}_{\overline
\zeta}q_{1}-M_{1})}{\partial_\zeta\Phi}\right
)\frac{e^{\tau(\Phi-\overline\Phi)}}{(\overline\zeta-\overline z)}
d\xi_2d\xi_1\nonumber\\-\frac{i}{2\pi}
\frac{\partial\psi}{\partial\nu}\int_{\Omega}
\partial_{\overline\zeta}\left (\frac{\overline{a}(\partial^{-1}_{\overline
\zeta}q_{1}-M_{3})}{\partial_{\overline\zeta}\overline\Phi}\right)\frac{e^{\tau(\overline\Phi-\Phi)}}{(\zeta-z)}
d\xi_2d\xi_1\quad\mbox{on}\,\,\Gamma_0^*
\end{eqnarray}
and
\begin{equation}\label{leopard}
\Vert a_{1,3}(\cdot,\tau)\Vert_{L^2(\Omega)}+\Vert
b_{1,3}(\cdot,\tau)\Vert_{L^2(\Omega)}=o(1)
\quad\mbox{as}\,\,\tau\rightarrow +\infty.
\end{equation}

The inequality (\ref{leopard}) follows from the asymptotic formula
\begin{eqnarray}\label{zopa}
\left\Vert \int_{\Omega}
\partial_\zeta\left (\frac{ a(\partial^{-1}_{\overline
\zeta}q_{1}-M_{1})}{\partial_\zeta\Phi}\right
)\frac{e^{\tau(\Phi-\overline\Phi)}}{\overline\zeta-\overline z}
d\xi_2d\xi_1\right\Vert_{W^{\frac 12}_2(\Gamma_0^*)} \nonumber\\
+ \left\Vert \int_{\Omega}
\partial_{\overline\zeta}\left (\frac{\overline{a}(\partial^{-1}_{\overline
\zeta}q_{1}-M_{3})}{\partial_{\overline\zeta}\overline\Phi}\right)
\frac{e^{\tau(\overline\Phi-\Phi)}}{\zeta-z}
d\xi_2d\xi_1\right\Vert_{W^{\frac 12}_2(\Gamma_0^*)}
=o(1)\quad\mbox{as}\,\,\tau\rightarrow +\infty.
\end{eqnarray}

In order to prove (\ref{zopa}) consider the function  $e\in
C^\infty_0(\Omega)$ such that $e\equiv 1$ in some neighborhood of
the set $\mathcal H\setminus\Gamma_0^*.$ The family of functions
$\int_{\Omega} e
\partial_\zeta\left (\frac{ a(\partial^{-1}_{\overline
\zeta}q_{1}-M_{1})}{\partial_\zeta\Phi}\right
)\frac{e^{\tau(\Phi-\overline\Phi)}}{\overline\zeta-\overline z}
d\xi_2d\xi_1\in C^\infty(\partial\Omega),$  are uniformly bounded in
$\tau$  in $C^2(\partial\Omega)$ and by Proposition 2.4 of \cite
{IUY} this function converges pointwisely to zero. Therefore
\begin{equation}\label{zopa1}
\left\Vert\int_{\Omega}e
\partial_\zeta\left (\frac{ a(\partial^{-1}_{\overline
\zeta}q_{1}-M_{1})}{\partial_\zeta\Phi}\right
)\frac{e^{\tau(\Phi-\overline\Phi)}}{\overline\zeta-\overline z}
d\xi_2d\xi_1\right\Vert_{W^1_2(\partial\Omega)}
=o(1)\quad\mbox{as}\,\,\tau\rightarrow +\infty.
\end{equation}

Integrating by parts we obtain
$$
\int_{\Omega}(1-e)
\partial_\zeta\left (\frac{ a(\partial^{-1}_{\overline
\zeta}q_{1}-M_{1})}{\partial_\zeta\Phi}\right
)\frac{e^{\tau(\Phi-\overline\Phi)}}{\overline\zeta-\overline z}
d\xi_2d\xi_1 =
\frac{(1-e)}{\partial_z\Phi}
\partial_z\left (\frac{ a(\partial^{-1}_{\overline z}q_{1}-M_{1})}
{\tau\partial_z\Phi}\right)e^{\tau(\Phi-\overline\Phi)}
$$
$$
-\frac{1}{\tau}\int_{\Omega}\partial_{\zeta}\left(\frac{(1-e)}
{\partial_\zeta\Phi}\partial_\zeta\left (\frac{ a(\partial^{-1}_{\overline
\zeta}q_{1}-M_{1})}{\partial_\zeta\Phi}\right )\right)
\frac{e^{\tau(\Phi-\overline\Phi)}}
{\overline\zeta-\overline z}d\xi_2d\xi_1 .
$$

Thanks to (\ref{1}) and (\ref{LL}), we have
\begin{equation}\label{zopa2}
\left\Vert \frac{1-e}{\partial_z\Phi}
\partial_z\left (\frac{ a(\partial^{-1}_{\overline z}q_{1}-M_{1})}
{\tau\partial_z\Phi}\right
)e^{\tau(\Phi-\overline\Phi)}\right\Vert_{W^{\frac
12}_2(\Gamma_0^*)}=o(1)\quad\mbox{as}\,\,\tau\rightarrow +\infty.
\end{equation}

The functions $\partial_{\zeta}\left(\frac{1-e}{\partial_\zeta\Phi}
\partial_\zeta\left (\frac{ a(\partial^{-1}_{\overline
\zeta}q_{1}-M_{1})}{\partial_\zeta\Phi}\right
)\right)e^{\tau(\Phi-\overline\Phi)}$ are bounded in $L^p(\Omega)$
uniformly in $\tau.$  Therefore by Proposition 2.2 of \cite{IUY},
the functions
$\int_{\Omega}\partial_{\zeta}\left(\frac{1-e}{\partial_\zeta\Phi}
\partial_\zeta\left (\frac{ a(\partial^{-1}_{\overline
\zeta}q_{1}-M_{1})}{\partial_\zeta\Phi}\right
)\frac{e^{\tau(\Phi-\overline\Phi)}}{\overline\zeta-\overline z}\right)
d\xi_2d\xi_1$ are uniformly bounded in $W^1_p(\Omega).$
The trace theorem yields
\begin{equation}\label{zopa3}
\left\Vert\frac{1}{\tau}\int_{\Omega}\partial_{\zeta}\left(
\frac{1-e}{\partial_\zeta\Phi}
\partial_\zeta\left (\frac{ a(\partial^{-1}_{\overline
\zeta}q_{1}-M_{1})}{\partial_\zeta\Phi}\right
)\frac{e^{\tau(\Phi-\overline\Phi)}}
{\overline\zeta-\overline z}\right) d\xi_2d\xi_1\right\Vert
_{W^{\frac 12}_2(\Gamma_0^*)}=o(1)\quad
\mbox{as}\,\,\tau\rightarrow +\infty.
\end{equation}
By (\ref{zopa1})-(\ref{zopa3}) we obtain (\ref{zopa}).

We note that by (\ref{LL}) the function $\frac{a}{\partial_z\Phi}\in
C^2(\partial\Omega).$
We define the function $U_1$ by the formula
\begin{eqnarray}\label{mozilaal}
U_1(x)=e^{\tau{\Phi}}{(a+a_{1}/\tau)} +e^{\tau\overline{\Phi}}
{(\overline a+b_{1}/\tau)}
- \frac 12e^{\tau\Phi}\widetilde {\mathcal
R}_\tau\{{a(\partial^{-1}_{\overline z} q_{1}-M_{1})}\} - \frac 12
 e^{\tau\overline\Phi}{\mathcal
R}_\tau\{\overline{a}(\partial^{-1}_{z} q_{1}-M_{3})\}.
\end{eqnarray}
Integrating by parts, we obtain the following:
\begin{equation}\label{01}
e^{\tau\Phi}\widetilde {\mathcal R}_\tau
\{{a(\partial^{-1}_{\overline z} q_{1}-M_{1})}\}
= \frac{1}{\tau}\Biggl(2b_{1,2}e^{\tau\overline\Phi}
+ \frac{e^{\tau\Phi}a(\partial^{-1}_{\overline z}
q_{1}-M_{1})}{2\partial_z\Phi}
\end{equation}
$$
+ \frac{e^{\tau\overline\Phi}}{2\pi}\int_{\Omega}
\partial_\zeta\left (\frac{ a(\partial^{-1}_{\overline
\zeta}q_{1}-M_{1})}{\partial_\zeta\Phi}\right
)\frac{e^{\tau(\Phi-\overline\Phi)}}{\overline\zeta-\overline z}
d\xi_2d\xi_1\Biggr)
$$
and
\begin{equation}\label{022}
e^{\tau\overline\Phi}{\mathcal R}_\tau\{\overline{a}(\partial^{-1}_{z}
q_{1}-M_{3})\}
= \frac{1}{\tau}\Biggl(2a_{1,2}e^{\tau\Phi}+\frac
{e^{\tau\overline\Phi}\overline{a}(\partial^{-1}_{z}
q_{1}-M_{3})}{2\partial_{\overline
z}\overline\Phi}
\end{equation}
$$
+ \frac{e^{\tau\Phi}}{2\pi}\int_{\Omega}
\partial_{\overline\zeta}\left (\frac{\overline{a}(\partial^{-1}_{\overline
\zeta}q_{1}-M_{3})}{\partial_{\overline\zeta}\overline\Phi}\right)
\frac{e^{\tau(\overline\Phi-\Phi)}}{\zeta-z} d\xi_2d\xi_1\Biggr).
$$

We claim that
\begin{eqnarray}\label{PPPP}
\left\Vert\frac{e^{-i\tau\psi}}{2\pi}\int_{\Omega}
\partial_\zeta\left (\frac{ a(\partial^{-1}_{\overline
\zeta}q_{1}-M_{1})}{\partial_\zeta\Phi}\right
)\frac{e^{\tau(\Phi-\overline\Phi)}}{\overline\zeta-\overline z}
d\xi_2d\xi_1\right\Vert_{L^2(\Omega)}\nonumber\\
+ \left\Vert
\frac{e^{i\tau\psi}}{2\pi}\int_{\partial\Omega}
\partial_{\overline\zeta}\left (\frac{\overline{a}(\partial^{-1}_{\overline
\zeta}q_{1}-M_{3})}{\partial_{\overline\zeta}\overline\Phi}
\right)\frac{e^{\tau(\overline\Phi-\Phi)}}{\zeta-z}
d\xi_2d\xi_1\right\Vert_{L^2(\Omega)} \rightarrow 0\quad
\mbox{as}\,\,\tau\rightarrow + \infty.
\end{eqnarray}
We prove the asymptotic formula (\ref{PPPP}) for the first term. The
proof of the asymptotic for the second term is the same. Denote
$r_\tau(\xi)=\partial_\zeta\left (\frac{ a(\partial^{-1}_{\overline
\zeta}q_{1}-M_{1})}{\partial_\zeta\Phi}\right
)e^{\tau(\Phi-\overline\Phi)}.$  By (\ref{22}), (\ref{begemot}) and
(\ref{begemot2}), the family of these functions is bounded in
$L^p(\Omega)$ for any $p<2.$ Hence by Proposition 2.2 of \cite{IUY}
there exists a constant $C$ independent of $\tau$ such that
\begin{equation}\label{mk}
\left\Vert\frac{e^{-i\tau\psi}}{2\pi}\int_{\Omega}
\partial_\zeta\left (\frac{ a(\partial^{-1}_{\overline
\zeta}q_{1}-M_{1})}{\partial_\zeta\Phi}\right
)\frac{e^{\tau(\Phi-\overline\Phi)}}{\overline\zeta-\overline z}
d\xi_2d\xi_1\right\Vert_{L^4(\Omega)}\le C.
\end{equation}

By (\ref{22}), (\ref{begemot}) and (\ref{begemot2}), for any $z\ne
\widetilde x_1+i\widetilde x_2$, the function $r_\tau(\xi)/(\bar
\zeta-\bar z)$ belongs to $L^1(\Omega).$ Therefore by Proposition
2.4 of \cite{IUY}, we have
\begin{equation}\label{mk1}
\frac{e^{-i\tau\psi}}{2\pi}\int_{\Omega}
\partial_\zeta\left (\frac{ a(\partial^{-1}_{\overline
\zeta}q_{1}-M_{1})}{\partial_\zeta\Phi}\right
)\frac{e^{\tau(\Phi-\overline\Phi)}}{\overline\zeta-\overline z}
d\xi_2d\xi_1\rightarrow 0\quad \mbox{a.e.  in }\quad \Omega.
\end{equation}

From (\ref{mk}), (\ref{mk1}) and Egorov's theorem, the asymptotic
for the first term in (\ref{PPPP}) follows immediately.

We set
$$
g_\tau=q_1(e^{i\tau\psi}{a_{1}/\tau} +e^{-i\tau\psi} {b_{1}/\tau}-
\frac{e^{i\tau\psi}}{2}\widetilde {\mathcal
R}_\tau\{{a(\partial^{-1}_{\overline z} q_{1}-M_{1})}\} -
 \frac{e^{-i\tau\psi}}{2}{\mathcal
R}_\tau\{\overline{a}(\partial^{-1}_{z} q_{1}-M_{3})\}).
$$
By (\ref{01})-(\ref{PPPP}) we have \begin{equation}
\label{inka}\Vert g_\tau\Vert_{L^2(\Omega)}=O(\frac 1\tau
)\quad\mbox{as}\,\,\tau\rightarrow +\infty.
\end{equation}
 Short computations give
\begin{equation}\label{lob}
L_1(x,D)U_1=e^{\tau\varphi}g_\tau\quad \mbox{in}\,\,\Omega,\quad
\frac{\partial
U_1}{\partial\nu}\vert_{\Gamma_0}=e^{\tau\varphi}O_{W_2^\frac
12(\overline{\Gamma^*_0})}(\frac
1\tau)\quad\mbox{as}\,\,\tau\rightarrow +\infty.
\end{equation}
Indeed, the first equation in (\ref{lob}) follows from
(\ref{mozilaal}), (\ref{begemot}) and the factorization of the
Laplace operator in the form $\Delta=4\partial_{\overline z}\partial_z.$
In order to prove the second equation in (\ref{lob}) we set
$\frac{\partial U_1}{\partial\nu}=I_1+I_2$ where
\begin{eqnarray}\label{luba}
I_1=\frac{\partial}{\partial\nu}((a+a_{1,1}/\tau)e^{\tau\Phi}+(\overline
a+b_{1,1}/\tau)e^{\tau\overline\Phi}) \nonumber\\
= e^{\tau\varphi}\left(i\tau\frac{\partial\psi}{\partial\nu}(a+a_{1,1}/\tau)
-i\tau\frac{\partial\psi}{\partial\nu}(\overline
a+b_{1,1}/\tau)+\frac{\partial}{\partial\nu}(a+a_{1,1}/\tau)
+ \frac{\partial}{\partial\nu}(\overline a+b_{1,1}/\tau)\right)
                                                        \nonumber\\
= \left(i\frac{\partial\psi}{\partial\nu}
\frac{a(\partial^{-1}_{\overline z}
q_{1}-M_{1})}{4\partial_z\Phi}-i\frac{\partial\psi}{\partial\nu}\frac
{\overline{a}(\partial^{-1}_{z} q_{1}-M_{3})}{4\partial_{\overline
z}\overline\Phi}\right)e^{\tau\varphi}
+ e^{\tau\varphi}O_{C^1(\overline\Gamma^*_0)}(\frac 1\tau).
\end{eqnarray}
In order to obtain the last equality, we used (\ref{LL}) and
(\ref{monica}). Then
\begin{eqnarray}\label{luba1}
I_2 = \frac{\partial}{\partial\nu}((a_{1,2}+a_{1,3})e^{\tau\Phi}
+(b_{1,2}+b_{1,3})e^{\tau\overline\Phi})
- \frac 12
\frac{\partial}{\partial\nu}(e^{\tau\Phi}\widetilde {\mathcal
R}_\tau\{{(a(\partial^{-1}_{\overline z}
q_{1}-M_{1})}\} \nonumber\\
+ e^{\tau\overline\Phi}{\mathcal
R}_\tau\{\overline{a}(\partial^{-1}_{z} q_{1}-M_{3})\})\nonumber\\
= -\frac 12\frac{\partial}{\partial\nu}\Biggl(
\frac{e^{\tau\Phi}a(\partial^{-1}_{\overline z}
q_{1}-M_{1})}{2\partial_z\Phi}+\frac{e^{\tau\overline\Phi}}{2\pi}
\int_{\Omega}
\partial_\zeta\left (\frac{ a(\partial^{-1}_{\overline
\zeta}q_{1}-M_{1})}{\partial_\zeta\Phi}\right
)\frac{e^{\tau(\Phi-\overline\Phi)}}{\overline\zeta-\overline z}
d\xi_2d\xi_1                         \nonumber\\
+ \frac{e^{\tau\overline\Phi}\overline{a}(\partial^{-1}_{z}
q_{1}-M_{3})}{2\partial_{\overline
z}\overline\Phi}+\frac{e^{\tau\Phi}}{2\pi}\int_{\Omega}
\partial_{\overline\zeta}\left (\frac{\overline{a}(\partial^{-1}_{\overline
\zeta}q_{1}-M_{3})}{\partial_{\overline\zeta}\overline\Phi}\right)
\frac{e^{\tau(\overline\Phi-\Phi)}}{\zeta-z} d\xi_2d\xi_1\Biggr)
                                                \nonumber\\
= -i\frac{\partial\psi}{\partial\nu}\left(\frac
{{a}(\partial^{-1}_{\overline z} q_{1}-M_{1})}{4\partial_{
z}\Phi}-\frac {e^{\tau\overline\Phi}\overline{a}(\partial^{-1}_{z}
q_{1}-M_{3})}{4\partial_{\overline z}\overline\Phi}\right) +
O_{W_2^{\frac 12}(\Gamma^*_0)}(\frac 1\tau).
\end{eqnarray}
From (\ref{luba}) and (\ref{luba1}), we obtain the second equation
in (\ref{lob}).

Finally we construct the last term of the complex geometric optics
solution $e^{\tau\varphi}w_\tau.$ Consider the boundary value
problem
\begin{equation}\label{lena}
L_{q_1}(x,D)(w_\tau e^{\tau\varphi})=-g_\tau
e^{\tau\varphi}\quad\mbox{in}\,\,\Omega,\quad \frac{\partial (w_\tau
e^{\tau\varphi})}{\partial\nu}\vert_{\Gamma_0}=-\frac{\partial
U_1}{\partial\nu}.
\end{equation}

By (\ref{inka}) and Proposition \ref{Theorem 2.3}, there exists a
solution to problem (\ref{lena}) such that
\begin{equation}\label{ioio}
\Vert w_\tau\Vert_{L^2(\Omega)}=o(\frac 1\tau) \quad
\mbox{as}\,\tau\rightarrow +\infty.
\end{equation}
Finally we set
\begin{equation}\label{ioioio}
u_1=U_1+e^{\tau\varphi} w_\tau.
\end{equation}
By (\ref{ioio}), (\ref{ioioio}), (\ref{leopard}) and
(\ref {mozilaal})-(\ref{022}), we can represent the complex geometric
optics solution $u_1$ in the form
\begin{eqnarray}\label{mozilaall}
u_1(x)=e^{\tau{\Phi}}{(a+(a_{1,1}+a_{1,2,1})/\tau)}
+e^{\tau\overline{\Phi}} {(\overline a+(b_{1,1}+b_{1,2,1})/\tau)}\nonumber\\
- \bigg( e^{\tau\Phi}\frac{a(\partial^{-1}_{\overline z}
q_{1}-M_{1})}{4\tau\partial_z\Phi} +
 e^{\tau\overline\Phi}\frac{\overline{a}(\partial^{-1}_{z}
q_{1}-M_{3})}{4\tau\overline{\partial_z\Phi}} \bigg
)+e^{\tau\varphi}o_{L^2(\Omega)}(\frac 1\tau)\quad
\mbox{as}\,\tau\rightarrow +\infty.
\end{eqnarray}

Since the Cauchy data (\ref{popo}) for the potentials $ q_1$ and
$q_2$ are equal, there exists a solution $u_2$  to the Schr\"odinger
equation with potential $q_2$ such that $\frac{\partial
u_1}{\partial \nu}=\frac{\partial u_2}{\partial\nu}$ on
$\partial\Omega$ and $ u_1= u_2$ on $\widetilde \Gamma$. Setting
$u=u_1-u_2$, we obtain
\begin{equation}\label{pp}
(\Delta+q_2)u=(q_2-q_1)u_1\quad \mbox{in}\,\,\Omega, \quad
u\vert_{\widetilde\Gamma}=\frac{\partial u}{\partial
\nu}\vert_{\partial\Omega}=0.
\end{equation}

In a similar way to the construction of $u_1$, we construct a complex
geometrical optics solution $v$ for the Schr\"odinger equation with
potential $q_2.$ The construction of $v$ repeats the
corresponding steps of the construction of $u_1.$ The only
difference is that instead of $q_{1}$ and $\tau$, we use $q_{2}$ and
$-\tau,$ respectively. We skip the details of the construction and
point out that similarly to (\ref{mozilaall}) it can be represented
in the form
\begin{eqnarray}\label{mozilaa}
v(x)=e^{-\tau{\Phi}}{(a+(\widetilde a_{1,1}+\widetilde
a_{1,2,1})/\tau)}
+e^{-\tau\overline{\Phi}} {(\overline a+(\widetilde b_{1,1}
+\widetilde b_{1,2,1})/\tau)}\nonumber\\
+ \left (e^{-\tau\Phi}\frac{a(\partial^{-1} _{\overline z}
q_{2}-M_{2})}{4\tau\partial_z\Phi} +e^{-\tau\overline\Phi}
\frac{\overline{a}(\partial^{-1}_{z} q_{2}-M_{4})}
{4\tau\overline{\partial_z\Phi}}\right
)+e^{-\tau\varphi}o_{L^2(\Omega)}(\frac 1\tau) \quad
\mbox{as}\,\tau\rightarrow +\infty,\quad \frac{\partial
v}{\partial\nu}\vert_{\Gamma_0}=0,
\end{eqnarray}
where $M_{2}(z)$ and $M_{4}(\overline z)$ satisfy
$$
(\partial^{-1}_{\overline z}q_{2} -M_{2})(\widetilde x)=0, \quad \quad
(\partial^{-1}_{z} q_{2} - M_{4})(\widetilde x)= 0.
$$
The functions $\widetilde a_1(z)=\widetilde a_{1,1}(z)+\widetilde
a_{1,2}(z)$ and $\widetilde b_1(z)=\widetilde b_{1,1}(z)+\widetilde
b_{1,2}(z)$ are given by
$$
-i\frac{\partial\psi}{\partial\nu} \widetilde a_{1,1}(z)
+i\frac{\partial\psi}{\partial\nu}
\widetilde b_{1,1}(\overline z)=-\frac{\partial (a+\overline a)}
{\partial\nu}+i\frac{\partial \psi}{\partial\nu}
\frac{a(\partial^{-1} _{\overline z}
q_{2}-M_{2})}{4\tau\partial_z\Phi} -i\frac{\partial
\psi}{\partial\nu} \frac{\overline{a}(\partial^{-1}_{z}
q_{2}-M_{4})} {4\tau\overline{\partial_z\Phi}}
\quad\mbox{on}\,\,\Gamma_0,
$$
\begin{equation}
\quad \widetilde a_{1,1}, \widetilde b_{1,1}\in
C^1(\overline\Omega)
\end{equation}
and $\widetilde a_{1,2,1}(z),
\widetilde b_{1,2,1}(\overline z)\in C^1(\overline \Omega)$
are holomorphic functions such that
$$
\widetilde b_{1,2,1}(\overline z)=\frac{1}{8\pi}\int_{\Gamma_0^*}
\frac{(\nu_1-i\nu_2) a(\partial^{-1}_{\overline
\zeta}q_{2}-M_{2})e^{\tau(\Phi-\overline\Phi)}}{(\overline\zeta-\overline
z)\partial_\zeta\Phi} d\sigma
$$
and
$$
\widetilde a_{1,2,1}(z)=\frac{1}{8\pi}\int_{\Gamma_0^*}
\frac{(\nu_1+i\nu_2)\overline{a}(\partial^{-1}_\zeta
q_{2}-M_{4})e^{\tau(\overline\Phi-\Phi)}}{(\zeta-z)\partial_{\overline\zeta}
\overline\Phi} d\sigma.
$$

Denote $q=q_1-q_2.$ Taking the scalar product of equation (\ref{pp})
with the function $v$, we have:
\begin{equation}
\int_\Omega q u_1vdx=0.
\end{equation}
From formulae (\ref{mozilaall}) and (\ref{mozilaa}) in the
construction of complex geometrical optics solutions, we have
\begin{eqnarray}\label{lala}
0=\int_\Omega q u_1vdx = \int_\Omega q(a^2+\overline
a^2)dx\nonumber\\
+ \frac 1\tau\int_\Omega q(
a(a_{1,1}+a_{1,2,1}+b_{1,1}+b_{1,2,1})+\overline{a}(\widetilde
a_{1,1}+\widetilde a_{1,2,1}+\widetilde b_{1,1}+\widetilde
b_{1,2,1}))dx                  \nonumber\\
+ \int_\Omega q(a\overline a e^{2\tau i\psi}+ a\overline a e^{-2\tau i\psi})dx
                                       \nonumber\\
+ \frac{1}{4\tau}\int_{\Omega} \left( qa^2 \frac{\partial_{\overline z}
^{-1}q_{2}-M_{2}} {\partial_z\Phi} + q\overline{a}^2
\frac{\partial_{z}^{-1}q_{2}
-{M_{4}}}{\overline{\partial_z\Phi}}\right)dx           \nonumber\\
- \frac{1}{4\tau}\int_\Omega\left( qa^2\frac{\partial_{\overline z}^{-1}
q_{1}-M_{1}}{\partial_z\Phi} +q\overline
a^2\frac{\partial_{z}^{-1}q_{1}-{
M_{3}}}{\overline{\partial_z\Phi}}\right)dx\nonumber\\
+ o(\frac{1}{\tau})=0\quad\mbox{as}\,\,\tau \rightarrow +\infty.
\end{eqnarray}

Since the potentials $q_j$ are not necessarily   from
$C^2(\overline\Omega)$, we can not directly use the stationary phase
argument (e.g., Evans \cite{E}). Let  function $\hat q\in
C^\infty_0(\Omega)$ satisfy $\hat q(\widetilde x)=q(\widetilde x).$
We have
\begin{equation}\label{rono}
\int_\Omega q\mbox{Re}\,(a\overline a e^{2\tau i\psi})dx
=\int_\Omega \hat q\mbox{Re}\,(a\overline a e^{2\tau
i\psi})dx+\int_\Omega (q-\hat q)\mbox{Re}\,(a\overline a e^{2\tau
i\psi})dx.
\end{equation}
Using the stationary phase argument and (\ref{begemot}), similarly
to \cite{IUY}, we obtain
\begin{equation}\label{masa}
\int_\Omega \hat q(a\overline a e^{2\tau i\psi} + a\overline a
e^{-2\tau i\psi})dx=\frac{2\pi (q\vert a\vert^2)(\widetilde
x)\mbox{Re}\,e^{2{\tau} i\psi(\widetilde x)}} {{\tau}
\vert(\mbox{det}\thinspace \psi'')(\widetilde x)\vert^\frac
12}+o\left(\frac{1}{{\tau}}\right)\quad\mbox{as}\,\,\tau \rightarrow
+\infty.
\end{equation}
For the second integral in (\ref{rono}) we obtain
$$
\int_\Omega (q-\hat q)(a\overline a e^{2\tau i\psi}+ a\overline a
e^{-2\tau i\psi})dx=\int_\Omega (q-\hat q)\left (a\overline a \frac
{(\nabla\psi,\nabla)e^{2\tau i\psi}}{2\tau i\vert
\nabla\psi\vert^2}-a\overline a  \frac {(\nabla\psi,\nabla)e^{-2\tau
i\psi}}{2\tau i\vert \nabla\psi\vert^2}\right )dx
$$
$$
= \int_{\partial\Omega} q\left (a\overline a \frac
{(\nabla\psi,\nu)e^{2\tau i\psi}}{2\tau i\vert
\nabla\psi\vert^2}-a\overline a  \frac {(\nabla\psi,\nu)e^{-2\tau
i\psi}}{2\tau i\vert \nabla\psi\vert^2}\right )d\sigma
$$
\begin{equation}\label{opl}-\frac{1}{2\tau i}
\int_\Omega\left \{ e^{2\tau i\psi}\mbox{div}\,\left ((q-\hat
q)a\overline a \frac {\nabla\psi}{\vert \nabla\psi\vert^2}\right )
- e^{-2\tau i\psi}\mbox{div}\,\left ( (q-\hat q
)a\overline a \frac{\nabla\psi}{\vert \nabla\psi\vert^2}\right )\right\}dx.
\end{equation}
Since $\psi\vert_{\Gamma_0}=0$ we have
$$
\int_{\partial\Omega} qa\overline a \left (\frac
{(\nabla\psi,\nu)e^{2\tau i\psi}}{2\tau i\vert \nabla\psi\vert^2}-
\frac {(\nabla\psi,\nu)e^{-2\tau i\psi}}{2\tau i\vert
\nabla\psi\vert^2}\right )d\sigma=\int_{\widetilde\Gamma}
\frac{qa\overline a}{2\tau i\vert \nabla\psi\vert^2}
(\nabla\psi,\nu)(e^{2\tau i\psi}- e^{-2\tau i\psi})d\sigma.
$$
By (\ref{1}), (\ref{kk}) and Proposition 2.4 in \cite{IUY} we
conclude that
\begin{equation}\nonumber
\int_{\partial\Omega} qa\overline a \left (\frac
{(\nabla\psi,\nu)e^{2\tau i\psi}}{2\tau i\vert \nabla\psi\vert^2}-
\frac {(\nabla\psi,\nu)e^{-2\tau i\psi}}{2\tau i\vert
\nabla\psi\vert^2}\right )d\sigma=o(\frac 1\tau)\quad
\mbox{as}\,\,\tau\rightarrow+\infty.
\end{equation}
The last integral over $\Omega$ in formula (\ref{opl}) is
$o(\frac{1}{\tau})$ and therefore
\begin{equation}\label{-3}
\int_\Omega(q-\hat q)(a\overline a e^{2\tau i\psi} + a\overline a
e^{-2\tau i\psi})dx=o(\frac 1\tau)\quad\mbox{as}\,\,\tau \rightarrow
+\infty.
\end{equation}
Taking into account that $\psi(\widetilde x)\ne 0$ and using (\ref{masa}),
(\ref{-3}) we have from (\ref{lala}) that
\begin{equation}
\frac{2\pi (q\vert a\vert^2)(\widetilde x)} {
\vert(\mbox{det}\thinspace \psi'')(\widetilde x)\vert^\frac 12}=0.
\end{equation}
Hence $q(\widetilde x)=0.$
 In \cite{IUY1} it is proved that there exists a holomorphic function $\Phi$
such that (\ref{1})-(\ref{kk}) are satisfied and a point $\widetilde
x\in \mathcal H$ can be chosen arbitrarily close to any given point
in $\Omega$ (see \cite{IUY}).  Hence we have $q\equiv 0.$ The proof
of the theorem is completed. $\square$

\end{document}